\documentclass{article}
\usepackage{graphicx}
\usepackage{amsmath}
\usepackage{amsfonts}

\begin{document}

\begin{center}
{\LARGE Spontaneous Symmetry Breaking through Mixing}\bigskip
\end{center}

\bigskip

\begin{center}
{\large Harald Fritzsch and Michael Spannowsky}

Ludwig Maximilian University

Sektion Physik, Theresienstr. 37, 80333 M\"{u}nchen Germany
\end{center}

\textbf{Abstract:}

We discuss a model, in which the negative mass square needed in the Higgs
mechanism is generated by mixing with a heavy scalar. We have two scalar
doublets in the standard model. Phenomenological properties of the heavy new
scalar are discussed. The heavy scalar can be detected by the LHC.

\bigskip

In the Standard Model (SM) the masses of the W- and Z-bosons and of the
fermions (leptons and quarks) are introduced by the mechanism of spontaneous
symmetry breaking ("Higgs mechanism"). It is assumed that the squared mass
term of the scalar particle is negative. This, together with the $\phi ^{4}$
self interaction of the scalar bosons leads to a non-zero vacuum expectation
value.

The masses of the W- and Z-bosons are spontaneously generated via the gauge
interaction, and the masses of the fermions appear due to their Yukawa-type
interactions with the scalar. No prediction can be made about the fermion
masses, since they depend on the Yukawa coupling constants, which are free
parameters.

The negative mass square of the scalar boson is rather peculiar and looks to
us like an ad-hoc assumption, not explaining why the electroweak gauge
symmetry is broken. In this paper we would like to propose a possible
underlying mechanism, able to generate the imaginary Higgs mass (see also 
\cite{Calmet}) at an energy scale testable at the LHC.

We assume that there is a massless scalar which mixes (seesaw like) with a
heavy scalar. Through the mixing a negative mass for the scalar is
introduced in a natural way. As a result of the mixing the spontaneous
symmetry breaking arises as in the Standard Model. For a long time, the
seesaw mechanism has been widely studied to generate light neutrino masses 
\cite{Mohap} as well as to explain the mass hierarchies in the whole fermion
sector \cite{Kiyo}. Instead, the idea of applying this mechanism to the
scalar sector of the SM beside incorporating new degrees of freedom is quite
new and recieved increasing interest over the last years \cite{Atwood}.

\bigskip

The scalar sector of this model is given by the following Lagrange density:%
\begin{equation}
L_{sc.}=D^{\mu }%
\begin{pmatrix}
\phi \\ 
\Phi%
\end{pmatrix}%
^{+}D_{\mu }%
\begin{pmatrix}
\phi \\ 
\Phi%
\end{pmatrix}%
-\left( \lambda \phi ^{+}\Phi +\lambda \Phi ^{+}\phi +\Lambda \Phi
^{2}-\kappa \left( \phi ^{+}\phi \right) ^{2}\right)
\end{equation}

The mass matrix is given by

\begin{equation}
M=%
\begin{pmatrix}
0 & \lambda \\ 
\lambda & \Lambda%
\end{pmatrix}%
,
\end{equation}

where we assume $\lambda \ll \Lambda $.

We diagonalize this mass matrix using%
\begin{equation}
S=%
\begin{pmatrix}
\cos \theta & -\sin \theta \\ 
\sin \theta & \cos \theta%
\end{pmatrix}%
.
\end{equation}

The mixing angle $\theta $ is fixed by $\lambda $ and $\Lambda $.

\begin{eqnarray}
M^{\prime } &=&SMS^{-1} \\
M^{\prime } &=&%
\begin{pmatrix}
1/2(\Lambda -\sqrt{4\lambda ^{2}+\Lambda ^{2}}) & 0 \\ 
0 & 1/2(\Lambda +\sqrt{4\lambda ^{2}+\Lambda ^{2}})%
\end{pmatrix}%
\end{eqnarray}

The assumption $\lambda \ll \Lambda $ ensures that $\left\vert \sin \theta
\right\vert \ll \left\vert \cos \theta \right\vert $.

Note that the negative mass squared for the light scalar is proportional to $%
\lambda ^{2}$ in first order.

\begin{eqnarray}
\mu ^{2}\left( \lambda ,\Lambda \right) &\equiv &-1/2(\Lambda -\sqrt{%
4\lambda ^{2}+\Lambda ^{2}})  \notag \\
\rho ^{2}\left( \lambda ,\Lambda \right) &\equiv &1/2(\Lambda +\sqrt{%
4\lambda ^{2}+\Lambda ^{2}})
\end{eqnarray}

where $\mu ^{2},\rho ^{2}>0$.

Beside the light scalar there is a heavy scalar $\Phi $, which is an isospin
doublet. As far as the light scalar is concerned, it acts like the Higgs
particle in the Standard Model. Besides this particle there is, however, the
new heavy scalar%
\begin{equation}
\Phi ^{\prime }=%
\begin{pmatrix}
H^{+} \\ 
H_{0}+iH_{1}%
\end{pmatrix}%
.
\end{equation}

with $\phi =\phi ^{\prime }\cos \theta -\Phi ^{\prime }\sin \theta $.

$S$ is, of course, an orthogonal transformation. Thus the purely kinetic
part of the lagrangian is not altered. There is no need to introduce for
this particle a self-interaction proportional to the field in the fourth
power or any Yukawa couplings. But because of the $\phi ^{4}$ term in the
scalar potential and the Yukawa coupling of $\phi $ there are couplings
between the $\Phi ^{\prime }$ and the $\phi ^{\prime }$ as well as Yukawa
like couplings for the heavy scalar, induced by the mixing. These new
interactions depend on the ratio $\lambda /\Lambda $. The larger the mass
difference between those particles, the more are non-standard-model
couplings suppressed in the Yukawa and scalar sector.

The minimum of the Higgs potential is given by 
\begin{equation}
\left\langle \phi \right\rangle =\sqrt{\frac{\mu ^{2}}{2\kappa \cos \theta }}
\end{equation}

and we have the tree-level relations%
\begin{equation}
M_{h}=\sqrt{2}\mu
\end{equation}

and%
\begin{equation}
M_{H}=\left( 3M_{h}^{2}\tan ^{2}\theta +2\rho ^{2}\right) ^{1/2}
\end{equation}

If the masses of the scalars were much heavier than the Higgs boson mass,
the mixing between those particles would be strongly suppressed and could be
neglected. For example, if the Higgs mass is $M_{h}=116~GeV$ and the new
scalars mass is $M_{\phi ^{\prime }}\approx 16~TeV$, these couplings are
suppressed by powers of the factor $\left\vert \sin \theta \right\vert
\approx 0.008$.

Neglecting the mixing terms it follows from $\left\langle \Phi ^{\prime
}\right\rangle =0$ that there are just four particle couplings between gauge
bosons and $\Phi ^{\prime }$, so the new scalars have to decay into
fermions. It decays almost exclusively to $t\bar{t}$ which makes it easily
possible to distinguish $\Phi ^{\prime }$ from $\phi ^{\prime }$
experimentally.

We would like to consider as an example the heavy scalar to have a mass
light enough to be produced at the LHC and the Higgs particle to have a mass
of about $140$ $GeV $. In this case the mixing cannot be neglected and the
scalars decay into fermions and - although still suppressed - into gauge
bosons.

The heavy neutral particle $H_{0}$ decays predominantly into fermionic
channels: 
\begin{equation*}
H_{0}\rightarrow \bar{f}f
\end{equation*}

For the partial widths we find%
\begin{equation}
\Gamma \left( H_{0}\rightarrow \bar{f}f\right) =\tan ^{2}\theta \frac{G_{\mu
}N_{c}}{2\sqrt{2}\pi }M_{H_{0}}m_{f}^{2}\beta _{f}^{3}
\end{equation}

with $\beta =\left( 1-4m_{f}^{2}/M_{H_{0}}^{2}\right) ^{1/2}$ being the
velocity of the fermions and $N_{c}$ the color factor $N_{c}=3\left(
1\right) $ for quarks (leptons).

The heavy charged scalar $H^{\pm }$ decays as follows:

\begin{equation*}
H^{\pm }\rightarrow \bar{t}b.
\end{equation*}

This decay channel suffers from a large irreducible background%
\begin{equation*}
gg\rightarrow t\bar{t}g,~gq\rightarrow t\bar{t}q.
\end{equation*}

The partial width for $H^{\pm }$ can be calculated numerically (see \cite%
{Hahn}).

The decay widths of heavy scalars into gauge bosons are suppressed compard
to the Standard Model Higgs particle by the factor $\sqrt{2}\left(
M_{h}/M_{H_{0}}\right) ^{4}\tan ^{2}\theta .$ The partial widths for $H^{0}$
are given by%
\begin{equation}
\Gamma \left( H_{0}\rightarrow VV\right) =\frac{G_{\mu }}{16\pi }\left( 
\frac{M_{h}}{M_{H_{0}}}\right) ^{4}M_{H_{0}}^{3}\tan ^{2}\theta ~\delta _{V}%
\sqrt{1-4x}\left( 1-4x+12x^{2}\right) ,~x=\frac{M_{V}^{2}}{M_{H_{0}}^{2}}
\end{equation}

with $\delta _{W}=2$ and $\delta _{Z}=1$ \cite{Lee}.

The largest contributions for the production cross section of a very heavy
scalar are the gluon and vector boson fusion processes. The associative
production with $W$ or $Z$ bosons is at least two orders of magnitude
smaller for a scalar particle of more than $1~TeV$.

Reduced by a factor $\arctan ^{2}\theta \,$compared to a Higgs particle the
gluon fusion cross section for a single $H_{0}$ can be expressed in leading
order by \cite{Georgi}%
\begin{equation}
\sigma \left( gg\rightarrow H_{0}\right) =\tan ^{2}\theta \frac{G_{\mu
}\alpha _{s}^{2}}{144\pi }\left\vert \frac{3}{4}\sum_{q}A_{1/2}\left( \tau
_{Q}\right) \right\vert ^{2}
\end{equation}

with $\tau _{Q}=M_{H_{0}}^{2}/4m_{Q}^{2}$ and the form factor%
\begin{equation}
A_{1/2}\left( \tau _{Q}\right) =2\left[ \tau _{Q}-\frac{1}{4}\left( \tau
_{Q}-1\right) \left( \log \frac{1+\sqrt{1-\tau _{Q}^{-1}}}{1-\sqrt{1-\tau
_{Q}^{-1}}}-i\pi \right) ^{2}\right] .
\end{equation}

For calculation of the hadronic cross section $pp\rightarrow H_{0}$ we can
use ref. \cite{Spira}.

The cross section for the vector boson fusion production channel $%
qq\rightarrow V^{\ast }V^{\ast }qq\rightarrow H_{0}qq$ is suppressed by $%
\sqrt{2}\left( M_{h}/M_{H}\right) ^{4}\tan ^{2}\theta .$ To calculate the
hadronic cross section we use the program VV2H \cite{Spira2}.

The heavy quark associated production is a third production channel \cite%
{AssocProc}. But this channel is in the here considered mass region at least
two orders of magnitude smaller than the gluon fusion channel.

The only heavy scalar production channel which is not suffering from any
mixing induced suppression mechanism is the scalar pair production by vector
boson fusion. But this cross-section is far below $1$ $fb$ and thus not
large enough to enhance the total production cross section decisively \cite%
{Djouadi}.

There is also a fusion production process for the $H^{+}$.%
\begin{equation*}
u\bar{d}\rightarrow H^{+}
\end{equation*}

But this process should be even smaller because the coupling between $H^{+}$
and the quarks is proportional to the fermion masses. Numerical results are
shown in (Table \ref{Ergebnis}).

To distinguish the scalar $H_{0}$ from the pseudoscalar $H_{1}$ it is
possible to use the process 
\begin{equation*}
H_{0}/H_{1}\rightarrow t\bar{t}\rightarrow \left( W^{+}b\right) \left( W^{-}%
\bar{b}\right)
\end{equation*}%
analogously to the Higgs boson case \cite{parity}.

\begin{table}[tbph]
\par
\begin{center}
\begin{tabular}{|c||c|c|}
\hline
& $M_{H_{0}}=504$ & $M_{H_{0}}=800$ \\ \hline\hline
$\lambda $ & $3.5\cdot 10^{4}$ & $5.6\cdot 10^{4}$ \\ \hline
$\Lambda $ & $11.5\cdot 10^{4}$ & $31\cdot 10^{4}$ \\ \hline
$M_{h_{0}}$ & $140.10$ & $140.04$ \\ \hline
$\theta $ & $15.66$ & $9.93$ \\ \hline
$\Gamma \left( H_{0}\rightarrow \bar{t}t\right) $ & $1.74$ & $2.18$ \\ \hline
$\Gamma \left( H^{-}\rightarrow \bar{t}b\right) $ & $1.86$ & $1.36$ \\ \hline
$\Gamma \left( H_{0}\rightarrow W^{+}W^{-}\right) $ & $3.3\cdot 10^{-2}$ & $%
8.9\cdot 10^{-3}$ \\ \hline
$\Gamma \left( H^{-}\rightarrow W^{-}Z\right) $ & $8.0\cdot 10^{-4}$ & $%
2.1\cdot 10^{-4}$ \\ \hline
$\sigma \left( pp\rightarrow H_{0}\right) $ & $310$ & $13$ \\ \hline
$\sigma \left( pp\rightarrow H_{0}qq\right) $ & $5.3\cdot 10^{-1}$ & $%
1.1\cdot 10^{-2}$ \\ \hline
\end{tabular}%
\par
\end{center}
\caption{The hadronic cross sections for gluon fusion $\protect\sigma \left(
pp\rightarrow H_{0}\right) $ and for vector boson fusion $\protect\sigma %
\left( pp\rightarrow qqH_{0}\right) $ are in $fb$. The partial widths and
masses are in $GeV$. $\protect\alpha _{s}$ was evaluated at the partonic
center of mass energy.}
\label{Ergebnis}
\end{table}

The scalars decay almost exclusively into fermionic channels and so their
detectability suffers from a large irreducible background. Especially for
heavy scalar particles above $2$ $TeV$ for which the production cross
section is far below $1~fb$ it seems to be very peculiar to detect them.

But for scalars with a mass below $1~TeV$ there is a good chance to find
them at the LHC after collecting enough integrated luminosity.

\bigskip

We would like to thank D.N. Gao for valuable discussions.


\bigskip

\begin{thebibliography}{99}
\bibitem{Calmet} X. Calmet Eur. Phys. J. C, \textbf{28} (2003) 451-454 ;
X.Calmet and J.F. Oliver \textit{hep-ph/0606209 }(2006).

\bibitem{Mohap} R.N. Mohapatra and G. Senjanovic Phys. Rev. Lett., \textbf{44%
} (1980) 165; D. Chang, R.N. Mohapatra Phys. Rev.D \textbf{32} (1985) 1248.

\bibitem{Kiyo} Y. Kiyo, T. Morozumi, P. Parada, M.N. Rebelo and M. Tanimoto 
Prog.Theor.Phys.\textit{\ }\textbf{101}(1999) 671; Z.G. Berezhiani,
R.Rattazzi Phys. Lett.\textit{\ }B,\textit{\ }\textbf{279} (1992)%
124\textit{; }A. Davidson, K.C. Wali Phys.Rev.Lett., \textbf{60} (1988) 1813.

\bibitem{Atwood} D. Atwood, S. Bar-Shalom and A. Soni Eur. Phys. J. C, 
\textbf{45 }(2006) 219; S. Bar-Shalom, D.Atwood and A. Soni, PoS HEP2005
(2006) 358.

\bibitem{parity} M. Kramer,.J.H. Khun, M.L. Stong, P.M. Zerwas Z.Phys.%
\textbf{\ }C \textbf{64} (1994) 21\textbf{\ }

\bibitem{Hahn} T. Hahn, M. Perez-Victoria Comput. Phys. Commun. \textbf{118 }%
(1999) 153.

\bibitem{Lee} B.W. Lee, C. Quigg and H.B. Thacker Phys. Rev. D \textbf{16}
(1977) 1519.

\bibitem{Georgi} H. Georgi, S. Glashow, M. Machacek and D. Nanopoulos Phys.
Rev. Lett.\textit{\ }\textbf{40} (1978) 692

\bibitem{Spira} M. Spira Nucl.Instrum.Meth. A \textbf{389} (1997) 357\textbf{%
\ }

\bibitem{Spira2} K. Jakobs \textit{et al }Eur.Phys.J. C \textbf{32}, S2
(2004) 19.

\bibitem{Djouadi} A. Djouadi, W. Kilian, M. M\"{u}hlleitner and P.M. Zerwas
Eur. Phys. J. C.,\textit{\ }\textbf{45 }(1999)\textbf{\ }10

\bibitem{AssocProc} J.N. Ng and P. Zakarauskas, TRI-PP-82/85 (1982).
\end{thebibliography}
\end{document}